\documentclass[a4paper,11pt]{article}
\usepackage{pos}

\title{Pion electric polarizabilities from lattice QCD}
\author[a,b,c]{Xu~Feng}
\author[d,e]{Taku~Izubuchi}
\author*[e,f]{Luchang~Jin}
\author[g,h]{Maarten~Golterman}

\affiliation[a]{
Center for High Energy Physics, Peking University, Beijing 100871, China
}

\affiliation[b]{
Collaborative Innovation Center of Quantum Matter, Beijing 100871, China
}

\affiliation[c]{
School of Physics and State Key Laboratory of Nuclear Physics and Technology, Peking University, Beijing 100871, China
}

\affiliation[d]{
Physics Department, Brookhaven National Laboratory, Upton, New York 11973, USA
}

\affiliation[e]{
RIKEN-BNL Research Center, Brookhaven National Laboratory, Building 510, Upton, NY 11973, USA
}

\affiliation[f]{
Department of Physics, University of Connecticut, Storrs, CT 06269, USA
}

\affiliation[g]{
Department of Physics and Astronomy, San Francisco State University, San Francisco, CA 94132, USA
}

\affiliation[h]{
Department of Physics and IFAE-BIST, Universitat Autònoma de Barcelona E-08193 Bellaterra, Barcelona, Spain
}

\emailAdd{xu.feng@pku.edu.cn}
\emailAdd{ljin.luchang@gmail.com}
\emailAdd{maarten@sfsu.edu}

\abstract{
We report a first principle lattice calculation of the pion electric polarizability $\alpha_\pi$ at the physical pion mass. First, we derive the master formula, which relates the pion polarizabilities with the position space hadronic Compton tensor, $\langle \pi | J_\mu(x) J_\nu(0) | \pi \rangle$. The finite volume error of the master formula is exponentially suppressed by the spatial extent of the lattice. Then, the hadronic tensor is calculated using domain wall fermions (DWF) directly at physical pion mass. The gauge ensembles are generated by the RBC-UKQCD collaborations.
}

\FullConference{%
 The 38th International Symposium on Lattice Field Theory, LATTICE2021
  26th-30th July, 2021
  Zoom/Gather@Massachusetts Institute of Technology
}

\def \ra {\rangle}
\def \la {\langle}
\def \nn {\nonumber}
\def \ba {\begin{eqnarray}}
\def \ea {\end{eqnarray}}

\begin{document}

\maketitle

\section{\label{sec:intro}Introduction}

Polarizabilities describe the leading energy shift for
a neutral particle in a constant electric and magnetic field.
In Minkowski space-time convention, we have~\cite{Moinester:2019sew}:
\ba
H_\text{eff} =
- \frac{4\pi}{2} \alpha E^2
- \frac{4\pi}{2} \beta B^2
\quad\text{(Minkowski)}\ ,
\ea
where $\alpha$ is the electric and $\beta$ is the magnetic polarizability.
In Euclidean space-time, the convention for the electric field is different. Therefore, we have
\ba
H_\text{eff} =
\frac{4\pi}{2} \alpha E^2
- \frac{4\pi}{2} \beta B^2
\quad\text{(Euclidean)}\ .
\ea
In the following discussion, we will use the Euclidean space-time convention by default.
Based on this definition, the polarizabilities of neutral particles can be related to
the low energy behavior of the hadronic Compton tensor, $\la N | J_\mu(x) J_\nu(0) | N \ra$.
For a charged particle, polarizability can also be defined 
via the low energy behavior of the hadronic Compton tensor after subtracting the Born term contribution.
The electric current operator is defined as follow
\ba
J_\mu(x) = J_\mu(t_x, \vec x)
= e \Big(
  e_u \bar u \gamma_\mu u
+ e_d \bar d \gamma_\mu d
+ e_s \bar s \gamma_\mu s
\Big)\ ,
\ea
where $e_u = 2/3$, $e_d = e_s = -1/3$, and $\alpha_\text{QED} = e^2/(4\pi)\approx 1/137$.
The $\gamma_\mu$ matrices satisfy the anti-commutation relation: $\{\gamma_\mu, \gamma_\nu\} = 2 \delta_{\mu,\nu}$.

Two-loop calculations of the pion polarizabilities using Chiral perturbation theory
have been done for both
the charged pion~\cite{Burgi:1996mm,Burgi:1996qi,Gasser:2006qa}
and the neutral pion~\cite{Bellucci:1994eb,Gasser:2005ud}.
There are also some lattice calculations from first principle using the background field method~\cite{Niyazi:2021jrz,Bignell:2020dze,Ding:2020hxw}.
There are also attempts to use hadronic Compton tensor with small momentum transfer
to extract polarizablities~\cite{Burkardt:1994pw,Wilcox:2021rtt}.
Realistic lattice calculations along this direction are also underway.

In this work, we derive different position space formulas using the hadronic tensor
to obtain the pion electric polarizabilities.
We demonstrate these formulas allow efficient lattice calculations and will show some numerical results.
In particular, we emphasis the finite volume errors of the master formula obtained are
exponentially suppressed by the spatial lattice size $L$.

\section{\label{sec:formula}Formulation}

\subsection{\label{subsec:neutral}Neutral pion}

We start the derivation in finite volume to avoid the infinities from the infinite volume. However, we assume a periodic boundary box with very large volume (much larger than the real lattice size) so the finite volume effects can be neglected. We will analyze the finite volume effects of possible lattice calculations after we obtained the final expression.

Consider the zero momentum neutral pion correlation function ($t_\text{snk} \gg 0 \gg t_\text{src}$) in the presence of very smooth and slow varying external vector potential $ A_\mu(x) = A_\mu(t_x,\vec x)$,
which is only non-zero in the middle region of the pion correlation function.
We expect the neutral pion correlation function takes the following form according to the definition of the pion polarizabilities.

\ba
\la T \pi^0(t_\text{snk}) \pi^0(t_\text{src}) \ra_{A_\mu}
&=&
\la T \pi^0(t_\text{snk}) \pi^0(t_\text{src}) \ra
% \\
% &&
% \hspace{-1cm}
% \times
\exp\Big({-
\frac{4\pi}{2}\int_x  \frac{\alpha_{\pi^0} E^2(x) - \beta_{\pi^0} B^2(x)}{L^3}
}\Big)\ ,
% \nn
\ea
where $\pi^0(t)$ is a pion field with vanishing spatial momentum.
We can also calculate the correlation function as a perturbative expansion
in the vector potential $A_\mu$:
\ba
\label{pionpropAmu}
% &&\hspace{-0.5cm}
\la T \pi^0(t_\text{snk}) \pi^0(t_\text{src}) \ra_{A_\mu}
% \\
&=&
\la T \pi^0(t_\text{snk}) \pi^0(t_\text{src}) \ra
+
\la T \pi^0(t_\text{snk})
\int_{x} i A_\mu (x) J_\mu (x)
\pi^0(t_\text{src}) \ra
% \nn
\\
&&\hspace{0.5cm}
+\frac{1}{2}
\la T \pi^0(t_\text{snk})
\int_{x} i A_\mu (x) J_\mu (x)
\int_{y} i A_\nu (y) J_\nu (y)
\pi^0(t_\text{src}) \ra\ .
\nn
\ea
For the neutral pion,
\ba
\la \pi^0(t_\text{snk}) J_\mu (x) \pi^0(t_\text{src}) \ra = 0\ .
\ea
Combine the above equations and use translational invariance, we obtain:
\ba
&&\hspace{-1.5cm}
\frac{4\pi}{2}\int_x  \frac{\alpha_{\pi^0} E^2(x) - \beta_{\pi^0} B^2(x)}{L^3}
\nn\\
&=&
\frac{1}{2}
\int_{x}
\Bigg(
\int_{y} 
A_\mu (x+y) A_\nu (y)
\Bigg)
\frac
{\la T \pi^0(t_\text{snk})
J_\mu (x) J_\nu (0)
\pi^0(t_\text{src}) \ra}
{\la T \pi^0(t_\text{snk}) \pi^0(t_\text{src})\ra}
\\
\label{eq:general}
&=&
\frac{1}{2}
\int_{x}
\Bigg(
\int_{y} 
A_\mu (x+y) A_\nu (y)
\Bigg)
\frac{1}{2 M_\pi}
\la \pi^0 |T J_\mu(x) J_\nu(0) | \pi^0 \ra\ .
\ea
In the second step,
we have rewrite the correlation function in terms of
the matrix elements of the zero momentum pion state.
Actually, we need to subtract the vacuum contribution in the above formula (note the above derivation is valid without the pion as well, in which case we would obtain the vacuum fermion sea energy shift due to the E\&M field):
\ba
\label{eq:vac-sub}
&&\hspace{-0cm}
\frac
{\la T \pi^0(t_\text{snk})
J_\mu (x) J_\nu (0)
\pi^0(t_\text{src}) \ra}
{\la T \pi^0(t_\text{snk}) \pi^0(t_\text{src})\ra}
\\
&\to&
\frac
{\la T \pi^0(t_\text{snk})
J_\mu (x) J_\nu (0)
\pi^0(t_\text{src}) \ra}
{\la T \pi^0(t_\text{snk}) \pi^0(t_\text{src})\ra} - \la T J_\mu(x) J_\nu(0)\ra\ .
\nn
\ea
We will assume this subtraction in later discussion without explicitly writing it down.
% The correlation function above can be written in terms of the matrix elements of the zero momentum pion state:
% \ba
% \frac{4\pi}{2}\int_x (\alpha_{\pi^0} E^2(x) - \beta_{\pi^0} B^2(x))
% &=&
% \frac{1}{2}
% \int_{x}
% \Bigg(
% \int_{y} 
% A_\mu (x+y) A_\nu (y)
% \Bigg)
% \nn\\
% &&\hspace{-1.5cm}
% \times
% \frac{1}{2 M_\pi}
% \la \pi^0 |T J_\mu(x) J_\nu(0) | \pi^0 \ra
% \ea

At this point, we can consider some specific choice of vector potential.
Focusing on the electric polarizability, we can choose:
\ba
\vec A(t_x)=\vec A (t_x,\vec x)=\vec A(x)\ ,
\ea
while taking the time component of the vector potential equal to zero.
Only the electric field is non-zero with this choice.
Apply the general formula Eq.~(\ref{eq:general}), we obtain:
\ba
\label{eq:ieff-alpha}
I_\text{eff}
&=&
\int_{t_x}
\frac{4\pi}{2} \alpha_{\pi^0} E^2(t_x)
\\
&=&
\int_{t_x,\vec x}
\Bigg(
\int_{t_y} 
A_i (t_x + t_y) A_j (t_y)
\Bigg)
% \nn\\&&\hspace{1cm}\times
\label{eq:ieff-mat}
\frac{1}{2 M_\pi}\frac{1}{2}
\la \pi^0 |T J_i(t_x,\vec x) J_j(0,\vec 0) | \pi^0 \ra\ .
\ea
Since we choose the vector potential to vary very slowly,
we can expand the time dependence of $\vec A$:
\ba
\label{eq:a-expansion}
A_i (t_x + t_y)\approx
A_i (t_y) + t_x \partial_t A_i (t_y)
+ \frac{1}{2} t_x^2 \partial_t^2 A_i (t_y)+\dots\ .
\ea
Due to the current conservation ($\partial^x_\mu J_\mu(t_x, \vec x) = 0$):
\ba
\int_{t_x,\vec x}
\la \pi^0 |T J_i(t_x,\vec x) J_j(0,\vec 0) | \pi^0 \ra
=
\int_{t_x,\vec x}
\partial^x_\mu
\la \pi^0 |T x_i J_\mu(t_x,\vec x) J_j(0,\vec 0) | \pi^0 \ra
=
0\ ,
\ea
and time reflection symmetry:
\ba
\int_{t_x,\vec x}
t_x 
\la \pi^0 |T J_i(t_x,\vec x) J_j(0,\vec 0) | \pi^0 \ra
=
0\ ,
\ea
the first two terms in Eq.~(\ref{eq:a-expansion}) do not contribute to $I_\text{eff}$
in Eq.~(\ref{eq:ieff-mat}).
Only the third term remains.
Therefore
\ba
I_\text{eff}
&=&
\frac{1}{2}
\int_{t_x,\vec x,t_y}
A_j (t_y)
\frac{1}{2} t_x^2 \partial_t^2 A_i (t_y)
% \nn\\&&\hspace{1cm}\times
\frac{1}{2 M_\pi} 
\la \pi^0 |T J_i(t_x,\vec x) J_j(0,\vec 0) | \pi^0 \ra\ .
\ea
Integrating by part and noticing the matrix element is only non-zero (after integration)
when $i = j$, we obtain:
\ba
I_\text{eff}
&=&
-
\frac{1}{2}
\int_{t_y}
\frac{1}{2} \partial_t \vec A (t_y)
\cdot  \partial_t \vec A (t_y)
\int_{t_x,\vec x}
\frac{1}{3}t_x^2
% \nn\\&&\hspace{1cm}\times
\frac{1}{2 M_\pi}
\la \pi^0 |T \vec J(t_x,\vec x) \cdot \vec J(0,\vec 0) | \pi^0 \ra\ .
\ea
Comparing the above equation with Eq.~(\ref{eq:ieff-alpha}),
we obtain for $\alpha_{\pi^0}$:
\ba
\label{eq:master}
\alpha_{\pi^0}
&=&
-\int_{t_x,\vec x}
\frac{t_x^2}{24\pi}
\frac{1}{2 M_\pi}\la \pi^0 |T \vec J(t_x,\vec x) \cdot \vec J(0,\vec 0) | \pi^0 \ra\ .
\ea
This is our master formula in this paper to obtain $\alpha_{\pi^0}$.
While the above equation seems positive definite,
it does not imply $\alpha_{\pi^0} > 0$.
The reason is that the vacuum contribution needs to be subtracted as indicated
in Eq.~(\ref{eq:vac-sub}).

It is possible to
use the general Eq.~(\ref{eq:general})
with a different choice of $A_\mu(x)$.
For example, we can choose a mostly time independent field $A_\mu(x) = A_\mu(\vec x)$
within a very long time interval with length $t_\text{int}$.
The fields smoothly vanishes outside the time interval.
We have the following relations:
\ba
\int_{t_x,\vec x}
E^2(t_x,\vec x)
&\approx&
t_\text{int}\int_{\vec x}
E^2(\vec x)\ ,
\\
\int_{t_x,\vec x}
B^2(t_x,\vec x)
&\approx&
t_\text{int}\int_{\vec x}
B^2(\vec x)\ .
\ea
With this choice of vector potential, we can similarly derive the
the following formulas for $\alpha_{\pi^0}$ and $\beta_{\pi^0}$:
\ba
\label{eq:other-1}
\alpha_{\pi^0} &=&
- \int_{t_x,\vec x}
\frac{\vec x^2}{24\pi}
\frac{1}{2 M_\pi}\la \pi^0 |T J_t(t_x,\vec x) J_t(0,\vec 0) | \pi^0 \ra\ ,
\\
\label{eq:other-2}
\beta_{\pi^0} &=&
\int_{t_x,\vec x}
\frac{ \vec x^2 }{48\pi}
\frac{1}{2 M_\pi}\la \pi^0 |T \vec J(t_x,\vec x) \cdot \vec J(0,\vec 0) | \pi^0 \ra
\\
\label{eq:other-3}
&=&
-\int_{t_x,\vec x}\frac{1}{24\pi}
\frac{1}{2 M_\pi}
\la \pi^0 |T \vec x \cdot \vec J(t_x,\vec x) \vec x \cdot \vec J(0,\vec 0) | \pi^0 \ra\ .
\nn
\ea
Since the matrix elements satisfy the current conservation condition,
we can obtain different but equivalent formulas
for both $\alpha_{\pi^0}$ and $\beta_{\pi^0}$.

\subsection{\label{subsec:neutral}Charged pion}
The charged pion polarizabilities are more conveniently defined via
low-energy expansion coefficients of the (virtual) Compton scattering processes
after the leading generalized Born terms being subtracted.
\ba
\label{eq:born-sub}
% &&\hspace{-1cm}
\la \pi |T J_\mu(t_x,\vec x) J_\nu(0,\vec 0) | \pi \ra_S
% \nn \\
& = & \la \pi |T J_\mu(t_x,\vec x) J_\nu(0,\vec 0) | \pi \ra
% \nn\\ &&\hspace{1cm}
- \la \pi |T J_\mu(t_x,\vec x) J_\nu(0,\vec 0) | \pi \ra_\text{Born}\ .
\ea
The Born term can be defined as Eq.~(115) in Ref.~\cite{Moinester:2019sew}.
For the Compton scattering processes,
in case where the photons are on-shell, the Born term is equal to the
scalar QED contribution.
If the photons are off-shell, the (generalized) Born term is defined to include
the electromagnetic form factor of the pions.
The Euclidean space-time expression for the Born term is:
\ba
% &&\hspace{-1cm}
T^\text{Born}_{\mu,\nu}(q_t,\vec q)
% \nn\\
&=&
\int_{t_x,\vec x}
e^{i q_t t_x - i \vec q \cdot \vec x }
\la \pi | T J_\mu(t_x,\vec x) J_\nu(0,\vec 0) | \pi \ra
\\
&=&
e^2 F^2(q_t^2 + \vec q^2)
\Big(
2 \delta_{\mu,\nu}
- \frac{(2p+q)_\mu(2p+q)_\nu}{(p+q)^2 + M_\pi^2}
% \nn\\&&\hspace{1cm}
- \frac{(2p-q)_\mu(2p-q)_\nu}{(p-q)^2 + M_\pi^2}
\Big)\ ,
\ea
where $p = (i M_\pi, \vec 0)$.

After the Born term subtraction,
the polarizabilies for both neutral and charged pions can be defined
in a uniform way as the low-energy expansion coefficients.
Therefore, the derived formulas 
Eqs.~(\ref{eq:master},\ref{eq:other-1},\ref{eq:other-2},\ref{eq:other-3})
is valid for the charged pions if we use the Born term subtracted
Compton tensor defined in Eq.~(\ref{eq:born-sub}).
\footnote{For neutral pions, the formulas are valid with or without
the Born term subtraction, i.e. $\alpha_{\pi^0}^\text{Born} = 0$.}

In particular, for our master equation, Eq.~(\ref{eq:master}),
we can explicitly calculate the Born term contribution,
which we need to subtract.
Based on the Born term definition, we have:
\ba
% &&\hspace{-1cm}
\frac{\partial^2}{\partial q_t ^2}
T^\text{Born}_{k,k}(q_t,\vec 0) \Bigg|_{q_t = 0}
% \nn \\
\label{eq:born-term1}
&=&
-\int_{t_x,\vec x}
t_x^2
\la \pi | T J_k(t_x,\vec x) J_k(0,\vec 0) | \pi \ra_\text{Born}
\\
\label{eq:born-term2}
&=&
\frac{\partial^2}{\partial q_t ^2}
\Big(
e^2 F^2(q_t^2) 2 \delta_{k,k}
\Big)\ .
\ea
For charged pion, we have $F_{\pi^\pm}(q^2) \approx 1 - r_\pi^2 q^2 / 6$,
where $r_\pi = 0.659(4)~\mathrm{fm}$~\cite{ParticleDataGroup:2020ssz},
is the $\pi^\pm$ charge radius.
Combining Eqs.~(\ref{eq:master},\ref{eq:born-sub},\ref{eq:born-term1},\ref{eq:born-term2}),
we obtain the expression for $\alpha_{\pi^\pm}$:
\ba
\label{eq:master-pm}
\alpha_{\pi^\pm}
&=&
-\int_{t_x,\vec x}
\frac{t_x^2}{24\pi}
\frac{1}{2 M_\pi}\la \pi^\pm |T \vec J(t_x,\vec x) \cdot \vec J(0,\vec 0) | \pi^\pm \ra
% \nn\\ &&\hspace{1cm}
-
\alpha_{\pi^\pm}^\text{Born}\ ,
\ea
where $ \alpha_{\pi^\pm}^\text{Born} = - \alpha_\text{QED} \frac{r_\pi^2}{3 M_\pi} = -14.94(18)\times 10^{-4}~\mathrm{fm}^3$.

\section{Lattice calculation}

On the lattice, we calculate the Euclidean space-time hadronic Compton tensor from the four-point function:
\ba
\label{eq:hadronic-matrix-elements}
{1 \over 2 M_\pi}
\la \pi |T J_\mu(t_x,\vec x) J_\nu(0,\vec 0) | \pi \ra
=L^3\frac{\langle \pi(t_x+t_\text{sep}) J_\mu(t_x,\vec x) J_\nu(0,\vec 0) \pi^\dagger(-t_\text{sep}) \rangle_L}
{\langle \pi(t_x+t_\text{sep}) \pi^\dagger(-t_\text{sep}) \rangle_L}\ ,
\ea
where we use Coulomb gauge fixed wall source for the pion operator $\pi(t)$,
and $t_\text{sep}$ is the time separation between the current operator and the pion interpolating operator.
The separation $t_\text{sep}$ is fixed for each ensemble
and set to be large enough ($\sim 1.5~\mathrm{fm}$) to ensure projection to the pion state.
Calculation of the hadronic Compton tensor is performed
on four ensembles generated by the RBC and UKQCD collaborations.
The names and attributes of the ensembles are shown in Table~\ref{tab:latinfo}.
The properties of these ensembles are studied in detail in Ref.~\cite{RBC:2014ntl}.

\begin{figure}[h]
\centering
\includegraphics[width=0.5\textwidth]{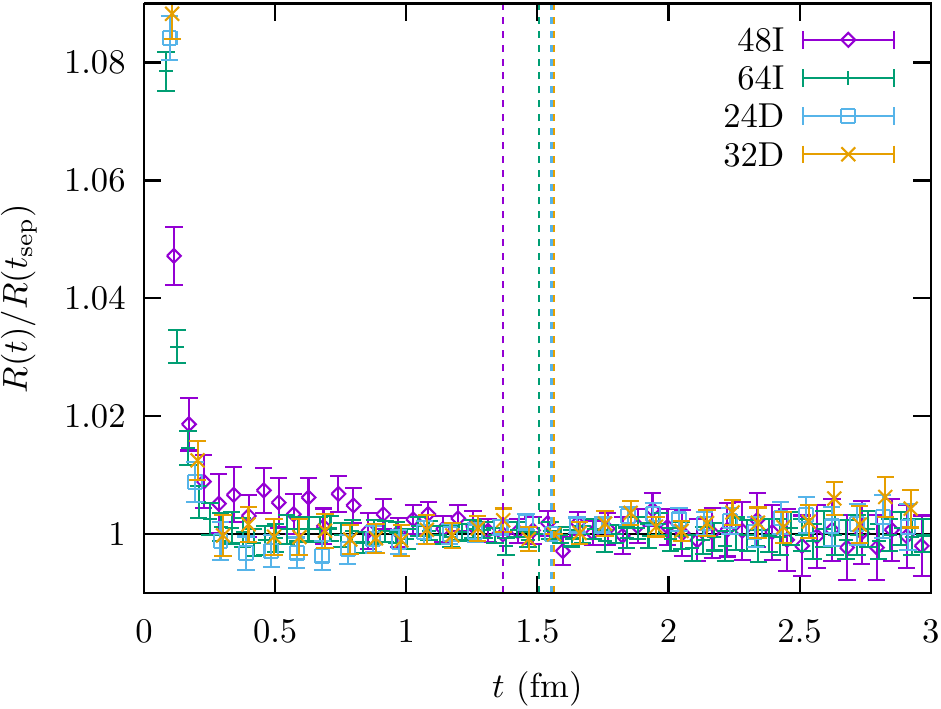}
\caption{
Ratio of the pion correlation function and its pion state contribution normalized at $t_\text{sep}$.
$R(t)$ is defined in Eq.~(\ref{eq:pion-corr-ratio}).
Vertical lines correspond to $t_\text{sep}$ used for each ensemble.
}
\label{fig:pion-corr-ratio}
\end{figure}

To study the size of the excited states contamination in our calculation,
we define the following ratio:
\ba
\label{eq:pion-corr-ratio}
R(t) = 
{
\langle \pi(t) \pi^\dagger(-t) \rangle_L
\over
\cosh\big( M_\pi (2\, t - T/2) \big)
}
\ea
The ratio should reach constant for large enough $t$, and its deviation from that
constant at small $t$ represent the norm of the excited states.
The lattice results of the above ratio is plotted in Figure~\ref{fig:pion-corr-ratio}.
It can be seen that the norms of the excited states are below 1 \% and are statistically insignificant.
\begin{table}[h]
\centering
\begin{tabular}{|c|c|c|c|c|c|}
\hline
 & Volume & $a^{-1}$ (GeV) & $L$ (fm) & $M_{\pi}$ (MeV) & $t_\text{sep}$ ($a$) \\ \hline
48I & $48^{3} \times 96$ & 1.730(4) & 5.5 & 135 & 12 \\ \hline
64I & $64^{3} \times 128$ & 2.359(7) & 5.4 & 135 & 18\\ \hline
24D & $24^3 \times 64$ & 1.0158(40) & 4.7 & 142 & 8\\ \hline
32D & $32^3 \times 64$ & 1.0158(40) & 6.2 & 142 & 8\\ \hline
% 32Dfine & $32^3 \times 64$ & 1.378(7) & 4.6 & 144 & 10 \\ \hline
% 24DH & $24^3 \times 64$ & 1.0158(40) & 4.7 & 341 & 8\\ \hline
\end{tabular}
\caption{
List of the ensembles used in the calculations and their properties.
They are generated by the RBC and UKQCD collaborations.~\cite{RBC:2014ntl}
Note we use a partially quenched quark mass for 48I and 64I ensembles.
The unitary pion mass for both 48I and 64I ensembles is $139~\mathrm{MeV}$.
We use unitary quark masses for all the other ensembles.
% Comment:
% Historically we made this choice, and the data were generated with this choice.
% For this project, I personally also think unitary pion mass would be slightly better.
}
\label{tab:latinfo}
\end{table}

\begin{figure}[h]
\centering
\includegraphics[width=0.3\textwidth]{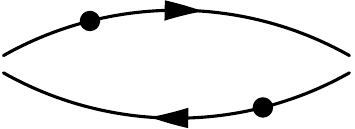}
\hspace{2cm}
\includegraphics[width=0.3\textwidth]{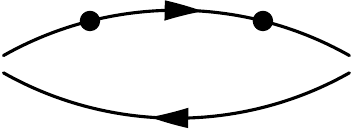}
\caption{
Quark connected diagrams of the four-point hadronic function (hadronic Compton tensor)
used to calculate the pion polarizabilities.
The dot represent the vector current operator insertion.
}
\label{fig:diagram}
\end{figure}
\begin{figure}[h]
\centering
\includegraphics[width=0.40\textwidth]{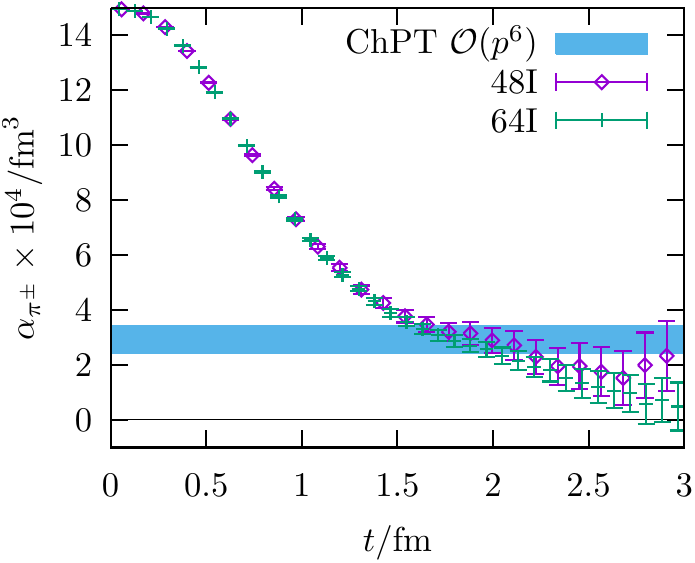}
~
\includegraphics[width=0.40\textwidth]{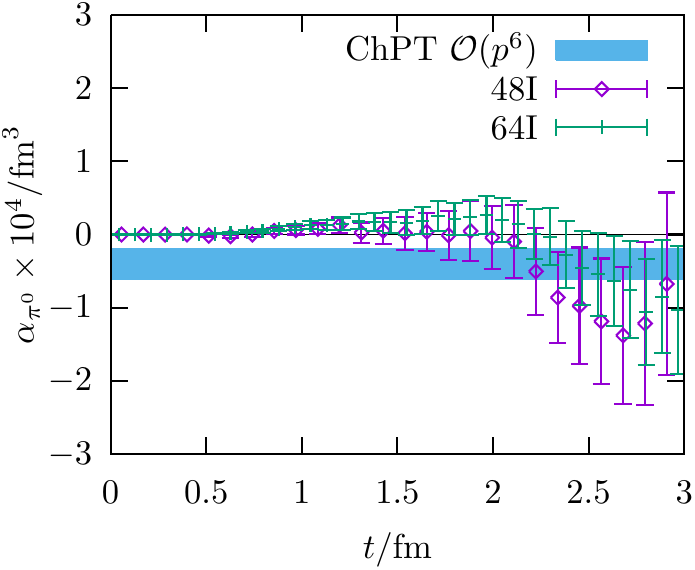}
\caption{
Electric polarizabilities of charged pion (left) and neutral pion (right).
The $t$ dependence is given by Eq.~(\ref{eq:master-partial-sum}), $\alpha_\pi = \alpha_\pi(t \to +\infty)$.
ChPT results are from Ref.~\cite{Burgi:1996mm,Burgi:1996qi,Gasser:2006qa,Bellucci:1994eb,Gasser:2005ud}.
}
\label{fig:master-results}
\end{figure}
\begin{figure}[h]
\centering
\includegraphics[width=0.40\textwidth]{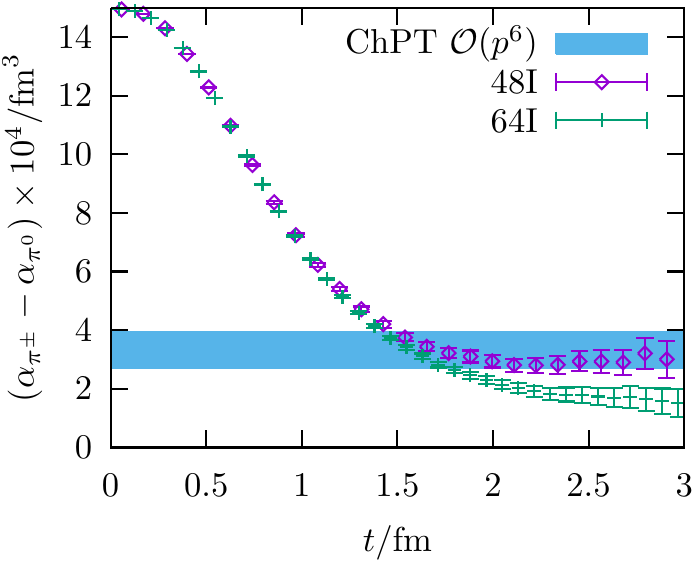}
~
\includegraphics[width=0.40\textwidth]{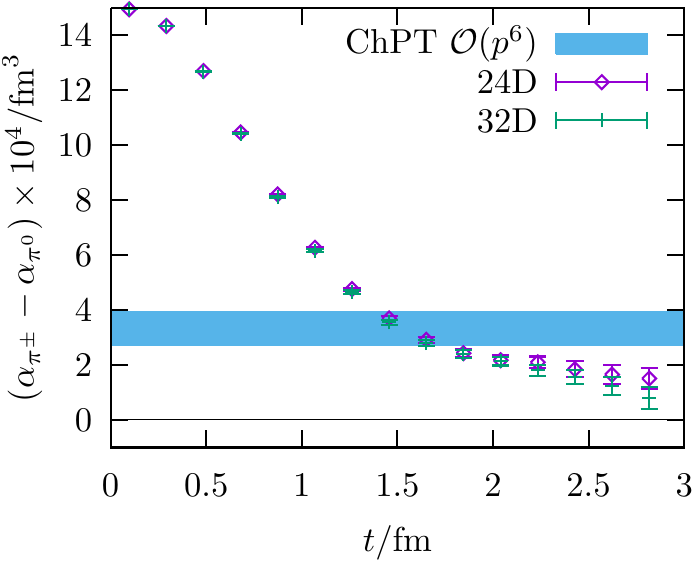}
\caption{
Difference of the electric polarizabilities between charged pion and neutral pion $\alpha_{\pi^\pm} - \alpha_{\pi^0}$.
The $t$ dependence is given by Eq.~(\ref{eq:master-partial-sum}), $\alpha_\pi = \alpha_\pi(t \to +\infty)$.
ChPT results are from Ref.~\cite{Burgi:1996mm,Burgi:1996qi,Gasser:2006qa,Bellucci:1994eb,Gasser:2005ud}.
}
\label{fig:master-results-diff}
\end{figure}

In this calculation of the pion polarizabilities, we only included the contribution
from the quark connected diagrams, which are shown in Figure~\ref{fig:diagram}.
We plot the results in Figure~\ref{fig:master-results} as a function of $t$:
\ba
\label{eq:master-partial-sum}
\alpha_\pi(t) = -
\int_{-t < t_x < t}\int_{\vec x}
\frac{t_x^2}{24\pi} \frac{1}{2 M_\pi}\la \pi |T \vec J(t_x,\vec x) \cdot \vec J(0,\vec 0) | \pi \ra
- \alpha_{\pi}^\text{Born}\ .
\ea
This is the partial sum of the $t$ integral of our master formula Eq.~(\ref{eq:master},\ref{eq:master-pm}),
and $\alpha_\pi = \alpha_\pi(t \to +\infty) $.

We notice that the difference between the electric polarizabilities of the charged pion
and the neutral pion $\alpha_{\pi^\pm} - \alpha_{\pi^0}$ only depends on the left
diagram in Figure~\ref{fig:diagram} (and a disconnected diagram which we ignore).
The numerical lattice result for this diagram is more precise than the other diagram.
Therefore, we plot the difference in Figure~\ref{fig:master-results-diff}.
In the right plot, we find the finite volume effect is very small
since the results from the 24D and 32D ensembles agree well.

\section{Conclusion}

We have derived several position space formulas for calculating
the polarizabilities of both neutral and charged hadrons on the lattice,
using pions as example.
The finite volume effect of all these type of formulas
is exponentially suppressed by the lattice size
and it is found to be numerically small
in our lattice calculation as well.
We also find the finite volume effects to be about 1\% in one-loop ChPT, for the charged minus neutral difference. 
We have calculated the electric polarizabilities for both charged and neutral pion.
% No significant deviation between our result and the ChPT based two-loop calculation is found.
At present, the lattice results tend to be lower than the ChPT predictions in the charged minus neutral difference.
Further investigation is needed to reach firm conclusions.
Improvement of the precision of the lattice calculation in the future is possible.

\section{Acknowledgements}
We would like to thank the RBC/UKQCD for the ensembles they have supplied and useful discussion many of the members therein have provided.
% We also think Prof. F.-K. Guo for the useful communications.
We also thank Heng-Tong Ding for helpful discussions, which initiate this work.
L.C.J. acknowledges support by DOE Office of Science Early Career Award DE-SC0021147 and DOE grant DE-SC0010339.
X.F. is supported in part by NSFC of China under Grants No. 11775002, No. 12125501 and No. 12070131001 and National Key Research and Development Program of China under Contracts No. 2020YFA0406400.
M.G. is supported by the U.S.Department of Energy, Office of Science, Office of High Energy Physics, under Award No. DE-SC0013682.  
We developed the computational code based on
the Columbia Physics System (https://github.com/RBC-UKQCD/CPS)
and Grid (https://github.com/paboyle/Grid).
The computation was performed under the ALCC Program of the US DOE on the Blue Gene/Q (BG/Q)
Mira computer at the Argonne Leadership Class Facility, a DOE Office of Science Facility
supported under Contract DE-AC02-06CH11357.
Computations for this work were carried out in part on facilities of the USQCD Collaboration, which are funded by the Office of Science of the U.S. Department of Energy.


\begin{thebibliography}{99}

%\cite{Moinester:2019sew}
\bibitem{Moinester:2019sew}
M.~Moinester and S.~Scherer,
%``Compton Scattering off Pions and Electromagnetic Polarizabilities,''
Int. J. Mod. Phys. A \textbf{34}, no.16, 1930008 (2019)
doi:10.1142/S0217751X19300084
[arXiv:1905.05640 [hep-ph]].
%6 citations counted in INSPIRE as of 24 Dec 2021

%\cite{Burgi:1996mm}
\bibitem{Burgi:1996mm}
U.~Burgi,
%``Charged pion polarizabilities to two loops,''
Phys. Lett. B \textbf{377}, 147-152 (1996)
doi:10.1016/0370-2693(96)00304-8
[arXiv:hep-ph/9602421 [hep-ph]].
%80 citations counted in INSPIRE as of 24 Dec 2021

%\cite{Burgi:1996qi}
\bibitem{Burgi:1996qi}
U.~Burgi,
%``Charged pion pair production and pion polarizabilities to two loops,''
Nucl. Phys. B \textbf{479}, 392-426 (1996)
doi:10.1016/0550-3213(96)00454-3
[arXiv:hep-ph/9602429 [hep-ph]].
%143 citations counted in INSPIRE as of 24 Dec 2021

%\cite{Gasser:2006qa}
\bibitem{Gasser:2006qa}
J.~Gasser, M.~A.~Ivanov and M.~E.~Sainio,
%``Revisiting gamma gamma ---\ensuremath{>} pi+ pi- at low energies,''
Nucl. Phys. B \textbf{745}, 84-108 (2006)
doi:10.1016/j.nuclphysb.2006.03.022
[arXiv:hep-ph/0602234 [hep-ph]].
%135 citations counted in INSPIRE as of 24 Dec 2021

%\cite{Bellucci:1994eb}
\bibitem{Bellucci:1994eb}
S.~Bellucci, J.~Gasser and M.~E.~Sainio,
%``Low-energy photon-photon collisions to two loop order,''
Nucl. Phys. B \textbf{423}, 80-122 (1994)
[erratum: Nucl. Phys. B \textbf{431}, 413-414 (1994)]
doi:10.1016/0550-3213(94)90566-5
[arXiv:hep-ph/9401206 [hep-ph]].
%190 citations counted in INSPIRE as of 24 Dec 2021

%\cite{Gasser:2005ud}
\bibitem{Gasser:2005ud}
J.~Gasser, M.~A.~Ivanov and M.~E.~Sainio,
%``Low-energy photon-photon collisions to two loops revisited,''
Nucl. Phys. B \textbf{728}, 31-54 (2005)
doi:10.1016/j.nuclphysb.2005.09.010
[arXiv:hep-ph/0506265 [hep-ph]].
%72 citations counted in INSPIRE as of 24 Dec 2021

%\cite{Niyazi:2021jrz}
\bibitem{Niyazi:2021jrz}
H.~Niyazi, A.~Alexandru, F.~X.~Lee and M.~Lujan,
%``Charged pion electric polarizability from lattice QCD,''
Phys. Rev. D \textbf{104}, no.1, 014510 (2021)
doi:10.1103/PhysRevD.104.014510
[arXiv:2105.06906 [hep-lat]].
%1 citations counted in INSPIRE as of 24 Dec 2021

%\cite{Bignell:2020dze}
\bibitem{Bignell:2020dze}
R.~Bignell, W.~Kamleh and D.~Leinweber,
%``Pion magnetic polarisability using the background field method,''
Phys. Lett. B \textbf{811}, 135853 (2020)
doi:10.1016/j.physletb.2020.135853
[arXiv:2005.10453 [hep-lat]].
%7 citations counted in INSPIRE as of 24 Dec 2021

%\cite{Ding:2020hxw}
\bibitem{Ding:2020hxw}
H.~T.~Ding, S.~T.~Li, A.~Tomiya, X.~D.~Wang and Y.~Zhang,
%``Chiral properties of (2+1)-flavor QCD in strong magnetic fields at zero temperature,''
Phys. Rev. D \textbf{104}, no.1, 014505 (2021)
doi:10.1103/PhysRevD.104.014505
[arXiv:2008.00493 [hep-lat]].
%24 citations counted in INSPIRE as of 24 Dec 2021

%\cite{Burkardt:1994pw}
\bibitem{Burkardt:1994pw}
M.~Burkardt, J.~M.~Grandy and J.~W.~Negele,
%``Calculation and interpretation of hadron correlation functions in lattice QCD,''
Annals Phys. \textbf{238}, 441-472 (1995)
doi:10.1006/aphy.1995.1026
[arXiv:hep-lat/9406009 [hep-lat]].
%32 citations counted in INSPIRE as of 24 Dec 2021

%\cite{Wilcox:2021rtt}
\bibitem{Wilcox:2021rtt}
W.~Wilcox and F.~X.~Lee,
%``Towards charged hadron polarizabilities from four-point functions in lattice QCD,''
Phys. Rev. D \textbf{104}, no.3, 034506 (2021)
doi:10.1103/PhysRevD.104.034506
[arXiv:2106.02557 [hep-lat]].
%3 citations counted in INSPIRE as of 24 Dec 2021

%\cite{RBC:2014ntl}
\bibitem{RBC:2014ntl}
T.~Blum \textit{et al.} [RBC and UKQCD],
%``Domain wall QCD with physical quark masses,''
Phys. Rev. D \textbf{93}, no.7, 074505 (2016)
doi:10.1103/PhysRevD.93.074505
[arXiv:1411.7017 [hep-lat]].
%274 citations counted in INSPIRE as of 27 Dec 2021

%\cite{ParticleDataGroup:2020ssz}
\bibitem{ParticleDataGroup:2020ssz}
P.~A.~Zyla \textit{et al.} [Particle Data Group],
%``Review of Particle Physics,''
PTEP \textbf{2020}, no.8, 083C01 (2020)
doi:10.1093/ptep/ptaa104
%2525 citations counted in INSPIRE as of 27 Dec 2021

\end{thebibliography}
\end{document}